\documentclass[preprint2]{aastex}
\usepackage{graphicx}
\usepackage{subfigure}

\shorttitle{Dark matter in massive early-type galaxies from the SDSS}

\begin{document}


\title{Projected central dark matter fractions and densities \\
in massive early-type galaxies from the Sloan Digital Sky Survey}


\author{C. Grillo\altaffilmark{1,2,3}}
\email{cgrillo@eso.org}


\altaffiltext{1}{Excellence Cluster Universe, Technische Universit\"at M\"unchen, Boltzmannstr. 2, D-85748, Garching bei M\"unchen, Germany}
\altaffiltext{2}{Max-Planck-Institut f\"ur extraterrestrische Physik, Giessenbachstr., D-85748, Garching bei M\"unchen, Germany}
\altaffiltext{3}{Universit\"ats-Sternwarte M\"unchen, Scheinerstr. 1, D-81679 M\"unchen, Germany}


\begin{abstract}

We investigate in massive early-type galaxies the variation of their two-dimensional central fraction of dark over total mass and dark matter density as a function of stellar mass, central stellar velocity dispersion, effective radius, and central surface stellar mass density. We use a sample of approximately $1.7 \times 10^{5}$ galaxies from the Sloan Digital Sky Survey Data Release Seven (SDSS DR7) at redshift smaller than 0.33. We apply conservative photometric and spectroscopic cuts on the SDSS DR7 and the MPA/JHU value-added galaxy catalogs, to select galaxies with physical properties similar to those of the lenses studied in the Sloan Lens ACS (SLACS) Survey. The values of the galaxy stellar and total mass projected inside a cylinder of radius equal to the effective radius are obtained, respectively, by fitting the SDSS multicolor photometry with stellar population synthesis models, under the assumption of a Chabrier stellar initial mass function (IMF), and adopting a one-component isothermal total mass model with effective velocity dispersion approximated by the central stellar velocity dispersion. The plausibility of an isothermal model to represent the galaxy total mass distribution is supported by independent gravitational lensing and stellar dynamical analyses performed in the lens subsample, which is found here to represent nicely the entire galaxy sample. We find that within the effective radius the stellar mass estimates differ from the total ones by only a relatively constant proportionality factor. In detail, we observe that the values of the projected fraction of dark over total mass and the logarithmic values of the central surface dark matter density (measured in $M_{\odot}\, \mathrm{kpc}^{-2}$) have almost Gaussian probability distribution functions, with median values of $0.64^{+0.08}_{-0.11}$ and $9.1^{+0.2}_{-0.2}$, respectively. We discuss the observed correlations between these quantities and other galaxy global parameters and show that our results disfavor an interpretation of the tilt of the Fundamental Plane (FP) in terms of differences in the galaxy dark matter content and give useful information on the possible variations of the galaxy stellar IMF and dark matter density profile. Finally, we provide some observational evidence on the likely significant contribution of dry minor mergers, feedback from active galactic nuclei, and/or coalescence of binary black holes on the formation and evolution of massive early-type galaxies.

\end{abstract}





\section{Introduction}

The properties of the Universe at cosmological scales ($\gtrsim$ 1 Mpc) have been satisfactorily reproduced within the currently favored hierarchical model of structure formation, i.e., the concordance $\Lambda$CDM (cold dark matter) model (e.g., \citealt{spr06}; \citealt{kom09}). Despite this success, at smaller scales ($\approx$ 1 kpc) a complete understanding of the dark and baryonic matter interplay, that ultimately determines the internal structure of galaxies, is still missing. Interestingly, not only the amount and distribution, but also the presence of dark matter in early-type galaxies has been controversial for a long time. The lack of suitable and easily interpreted kinematical tracers, such as H{\footnotesize I} in spirals, at large radii and the degeneracy between the mass distribution and the anisotropy of the stellar orbits (e.g., \citealt{ger93}; \citealt{lok03}) have made the detection of dark matter through dynamical studies difficult. Spectroscopic observations of the light emitted by stars have though provided the first useful constraints on the mass distribution within approximately two times the value of the effective radius $R_{e}$ in local early-type galaxies (e.g., \citealt{van91}; \citealt{ber94}; \citealt{car95}; \citealt{ger01}; \citealt{cap06}). In the framework of collisionless collapse, fully self-consistent two-component (luminous and dark matter) dynamical models have been developed (\citealt{ber92}) and applied (\citealt{sag92}) to fit the photometric and spectroscopic data of a sample of nearby galaxies, indicating positive evidence for the presence of dark matter. More recently, Schwarzschild (orbit-based) dynamical models have also enabled a decomposition of the total mass into luminous and dark in a sample of Coma early-type galaxies (\citealt{tho07}). 

Other kinematical tracers, like cold atomic hydrogen and warm ionized gas, have supported the picture of a dark matter component additional to the luminous one in early-type galaxies (e.g., \citealt{bus93}; \citealt{fra94}). Unfortunately, these gaseous sources are either rare or too restricted radially to trace dark matter straightforwardly in statistically significant samples of galaxies. A different diagnostics of the gravitational potential is the diffuse X-ray emission from the hot gas present in many early-type galaxies. The properties of the gas have allowed in some cases to estimate, under the hypotheses of quasi-hydrostatic equilibrium and spherical symmetry, mass-to-light ratios on the order of 100 on radial scales of about 100 kpc (e.g., \citealt{mus94}; \citealt{loe99}). Two difficulties affect studies of this kind, a theoretical and an observational one: the physics of the hot cooling gas is complex to model and the temperature profile of the gas may be measured with considerable uncertainty. A different class of tracers includes globular clusters and planetary nebulae. The discrete kinematical data obtained by observing these objects orbiting around some nearby early-type galaxies have provided an estimate of the gravitational field in these galaxies, confirming the presence of dark matter halos (e.g., \citealt{mou90}; \citealt{arn98}). All the cited analyses agree reasonably well in revealing flat circular velocity curves (i.e., a total matter density distribution well approximated by a $1/r^{2}$ profile) from $\gtrsim 0.2$ to $\gtrsim 2$ $R_{e}$ in early-type galaxies. Not only does the associated mass-density profile differ significantly from cosmologically motivated ones (e.g., \citealt{nav96}), but it also requires a significant amount of fine-tuning (known as bulge-halo ``conspiracy'') between the distribution of luminous and dark matter. This high degree of homology is still poorly understood in the currently accepted scenario of galaxy formation and evolution.

During the past decade, gravitational lensing has been extensively used to study the mass distribution of early-type galaxies beyond the local Universe, contributing to key observations of dark matter (e.g., \citealt{rus03}; \citealt{koo06}; \citealt{gri09}; \citealt{veg10}). The combination of strong lensing and stellar kinematics has proved to be particularly effective, since the two diagnostics complement each other (e.g., \citealt{tre04}; \citealt{czo08}; \citealt{bar09}). In fact, strong lensing provides a robust measurement of the total mass projected inside the Einstein radius (\citealt{koc91}), breaking the mass-anisotropy degeneracy of the dynamical analysis, after which stellar dynamics, giving constraints on the total mass distribution at small radii (typically $\lesssim R_{e}$), provides an estimate of the mass gradient. Thus, a joint strong lensing and stellar-dynamical study can be used to determine the average logarithmic density slope of the total mass inside the Einstein radius of individual lens galaxies, independent of the mass-sheet degeneracy (e.g., \citealt{fal85}; \citealt{koc06}). In addition, this allows one to decompose the total mass distribution into luminous and dark with good precision, making feasible an investigation on the evolution of the internal structure of early-type galaxies. This technique has been applied to a sample of 5 massive early-type galaxies in the redshift range $z \approx 0.5-1$, as part of the Lenses Structure and Dynamics (LSD) Survey (\citealt{tre04}), and to a larger sample of 58 massive early-type galaxies at lower redshifts ($z \approx 0.06-0.5$) from the Sloan Lens ACS (SLACS) Survey (\citealt{koo09}). Despite some scatter in individual galaxies, one of the most important results of these two surveys is the measurement in the complete sample of a remarkably homogeneous total (luminous and dark) mass density profile, that is consistent with an isothermal one (i.e., $\rho \propto 1/r^{2}$) out to a few hundreds of kiloparsecs (see also \citealt{gav07}; \citealt{bol08b}). Additional support to the picture of an average isothermal total mass density profile with a surprisingly small scatter has been provided by a fully self-consistent analysis of lensing data and two-dimensional kinematic maps in a smaller subsample (e.g., \citealt{czo08}; \citealt{bar09}). Moreover, no evidence of significant evolution in the value of the total mass density slope below redshift of 1 has been found. The measured one-component isothermal model is completely characterized by the value of an effective velocity dispersion, which has resulted to be nicely approximated (within less than 3\%) by the value of the galaxy central stellar velocity dispersion (i.e., the projected velocity dispersion of the stars within a disk of radius $R_{e}/8$; see, e.g., \citealt{tre06}; \citealt{bol08b}). This latter result is expected when considering the Jeans equation for realistic stellar density distributions (e.g., \citealt{jaf83}; \citealt{her90}) embedded in a globally isothermal distribution (see \citealt{koc93}) and some observational evidence has been collected by means of dynamical modeling in samples of nearby and luminous early-type galaxies (e.g., \citealt{koc94}; \citealt{gri08b}).

Strong lensing has also turned out to be an invaluable astrophysical tool when combined with stellar population synthesis models. Taking advantage of this combination, recent studies have addressed disparate topics that are relevant in the field of galaxy formation and evolution, such as the measurement of the projected dark over total mass fractions inside the inner regions of distant early-type galaxies (e.g., \citealt{gri08a,gri08c,gri09,gri10a}; \citealt{aug09}; \citealt{fad10}), the investigation of the most plausible stellar IMF (e.g., \citealt{gri08a,gri09,gri10b}; \citealt{tre10}) and origin of the tilt of the FP (\citealt{gri09,gri10b}) of massive early-type galaxies. Moreover, different analyses have revealed that the SLACS lens galaxies are a representative sample of their SDSS parent sample with regard to photometric, spectroscopic, and environmental properties (\citealt{bol06, bol08a}; \citealt{tre06, tre09}; \citealt{gri09}). This last point encourages to adopt here a different perspective and exploit, in the light of the latest lensing results, the wealth of information already accumulated by the SDSS to study the projected central dark over total mass fractions and dark matter densities of massive (lens and non lens) early-type galaxies.

The paper is organized as follows. In Sect. 2, we describe how the sample of massive early-type galaxies is selected from the SDSS DR7 and the relevant galaxy photometric and spectroscopic values are estimated. In Sect. 3, we outline the method used to measure the central dark over total mass fractions and dark matter densities. In Sect. 4, we present our results and explore possible correlations between the dark matter and several other physical quantities of the galaxies in the sample. In Sect. 5, we discuss and interpret the results of our analysis within different galaxy formation and evolution scenarios. Finally, in Sect. 6, we summarize and draw conclusions. The aperture mass values discussed in this work are intended two-dimensional mass measurements, i.e., projected along the line of sight. All length, mass, velocity, and surface mass density values considered in our study are expressed, respectively, in units of kpc, $M_{\odot}$, km s$^{-1}$, and $M_{\odot}\, \mathrm{kpc}^{-2}$. The logarithms assume a base of 10 and dimensionless arguments obtained by dividing the studied quantities with their corresponding measurement units. Following the general solution of \citet{yor66} to the linear least-squares problem, we determine all the best-fit correlation lines between two variables taking into account the uncertainties in both coordinates. Throughout this work, we assume $H_{0}=70$ km s$^{-1}$ Mpc$^{-1}$, $\Omega_{\mathrm{m}}=0.3$, and $\Omega_{\Lambda}=0.7$.

\section{The sample}

\begin{figure}[tb]
\centering
\includegraphics[width=0.22\textwidth]{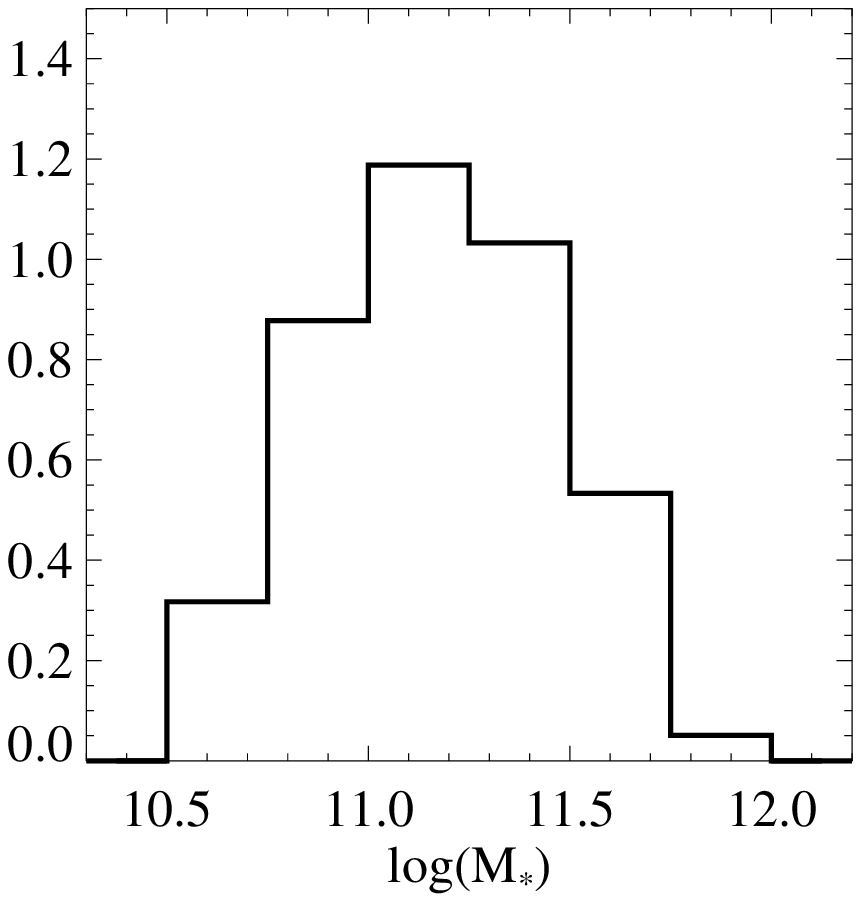}
\includegraphics[width=0.22\textwidth]{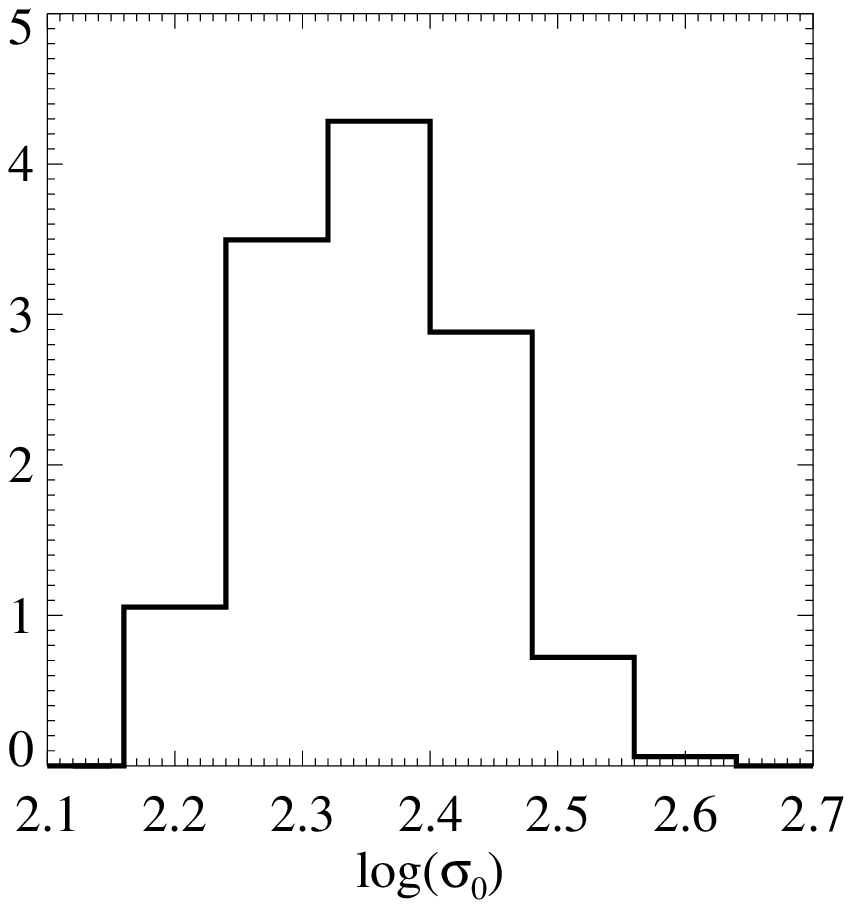}
\caption{The probability distribution functions of the logarithms of the total stellar mass $M_{*}$ (\emph{on the left}) and central stellar velocity dispersion $\sigma_{0}$ (\emph{on the right}) of the approximately $1.7 \times 10^{5}$ galaxies in the sample. The stellar masses and stellar velocity dispersions have been estimated, respectively, from the SDSS photometric and spectroscopic observations.}
\label{fi01}
\end{figure}

The sample of galaxies analyzed in this paper is drawn from the SDSS. This survey is providing multi-band photometry in five bandpasses (\emph{u}, \emph{g}, \emph{r}, \emph{i}, \emph{z}; see \citealt{fuk96}) for almost a quarter of the sky at high Galactic latitude and spectroscopic observations (taken using 3-arcsec diameter fibers; see \citealt{str02}) for more than $10^{6}$ objects. In addition to the data retrieved from the SDSS DR7 Galaxy catalog, we use here also the public galaxy catalogs provided by the MPA/JHU collaboration\footnote{http://www.mpa-garching.mpg.de/SDSS/}. 

First, we match the catalogs and select early-type galaxies by requiring that the surface brightness profile of each object is well described by a de Vaucouleurs profile. In detail, we make use of the SDSS morphological index \textsf{fracDeV}, which quantifies the weight of a de Vaucouleurs profile in a two-component (a de Vaucouleurs plus an exponential profile) decomposition  of the galaxy surface brightness distribution. We apply a highly conservative cut choosing only those objects that exhibit \emph{r}, \emph{i}, and \emph{z} \textsf{fracDeV} values greater than 0.95. Moreover, to make the sample more homogeneous we exclude the galaxies that belong to the subclasses called \textsf{AGN}, \textsf{BROADLINE}, and \textsf{STARFORMING} in the MPA/JHU galaxy catalog.

Then, we decide to concentrate our study only on massive early-type galaxies with physical scales similar to those of the lens galaxies studied in the SLACS survey (see \citealt{bol08a,gri09,aug09}). Accordingly, we select the galaxies that have SDSS spectroscopic redshifts $z_{\mathrm{sp}}$ between 0.05 and 0.5, aperture stellar velocity dispersions $\sigma_{\mathrm{ap}}$ that range from 150 and 400 km s$^{-1}$, and total stellar masses $M_{*}$ between $10^{10.5}$ and $10^{12}$ M$_{\odot}$. Stellar masses of the SDSS galaxies are measured by the MPA/JHU team by fitting the SDSS broad-band photometry (\textsf{modelMag} magnitudes) with a large grid of stellar population synthesis models, that can accommodate various star formation histories. \citet{bru03} templates are used and a \citet{cha03} stellar IMF is assumed (for more details on the stellar mass measurements, see \citealt{kau03}; \citealt{sal07}). In the catalog, the stellar mass estimates are available only for galaxies at redshift smaller than 0.33. Therefore, we restrict our sample to this upper limit in redshift.

We compute the rest frame \emph{r-}band effective angles $\theta_{e}$ (i.e., the angles within which half of the total light of a de Vaucouleurs profile is included) of the galaxies in the sample by linear interpolation in wavelength between the effective angles $\theta_{e,r}$, $\theta_{e,i}$, and $\theta_{e,z}$ that are fitted, considering the appropriate values of the point spread function, on the galaxy surface brightness distributions in the \emph{r-}, \emph{i-}, and \emph{z-}band, respectively. For these bands, we take their central wavelengths $\lambda_{r}$, $\lambda_{i}$, and $\lambda_{z}$ to be correspondingly 6231, 7625, and 9134 \AA, resulting in these expressions for the galaxy angular sizes:
\begin{eqnarray}
\theta_{e}&=\theta_{e,r} + \big[(1+z_{\mathrm{sp}})\lambda_{r}-\lambda_{r}\big] \frac{\theta_{e,i} - \theta_{e,r}}{\lambda_{i}-\lambda_{r}} \nonumber \\
\theta_{e}&=\theta_{e,i} + \big[(1+z_{\mathrm{sp}})\lambda_{r}-\lambda_{i}\big] \frac{\theta_{e,z} - \theta_{e,i}}{\lambda_{z}-\lambda_{i}}. \label{eq:01}
\end{eqnarray}
The first and second formulae shown in Eq. (\ref{eq:01}) are used, respectively, if the spectroscopic redshift of a galaxy is smaller or larger than 0.224. The effective radii $R_{e}$ are determined by multiplying the values of the effective angles by the angular diameter distances at the galaxy redshifts.

Next, we estimate the central velocity dispersions of the galaxy stellar component, $\sigma_{0}$, starting from the SDSS velocity dispersions, $\sigma_{\mathrm{ap}}$, that are measured within a 3-arcsec diameter fiber, and using the following empirical prescription determined by \citet{jor95}:
\begin{equation}
\sigma_{0} = \sigma_{\mathrm{ap}} \, \bigg( \frac{\theta_{e}}{8 \times 1.5 ''}\bigg)^{-0.04}.
\label{eq:02}
\end{equation}

The sample of massive early-type galaxies that we select according to the selection criteria described above consists of approximately $1.7 \times 10^{5}$ objects. In Fig. \ref{fi01}, we plot the resulting probability distribution functions of the logarithms of the mean values of the total stellar mass and of the values of the central stellar velocity dispersion. 

We emphasize that the total stellar mass estimates used in this study are robust. In fact, these quantities, that are measured from the SDSS multicolor photometry, agree very well (especially at the high mass end considered here) with the values determined by fitting different spectral indices in the galaxy spectra\footnote{A discussion on the observed differences between the estimates inferred from the photometric and spectroscopic stellar mass diagnostics is given at http://www.mpa-garching.mpg.de/SDSS/DR7/mass\_comp.html}. Moreover, in a recent study \citet{aug09} have compared in a subsample of approximately 50 massive lens galaxies the photometric stellar masses estimated by modeling in different ways (as done by the MPA/JHU collaboration and by \citealt{gri09}) the public SDSS photometry and their new HST photometry. Despite some minor differences, the authors conclude on the overall consistency of all the independently measured stellar masses. 

\section{The method}

We focus here on the determination of two-dimensional dark over total mass fractions and dark matter densities, measured within cylinders of radii $R_{e}$. We use the luminous (stellar) mass values introduced in the previous section and total mass values estimated from stellar dynamics. We intentionally concentrate on projected quantities for two reasons: first, because these quantities are more directly related to the observables, with no need of any deprojection modeling, and second, because these are the physical quantities that enter in several scaling laws of early-type galaxies, like the FP.

Based on the assumption that the light distribution of the galaxies traces well their stellar mass distribution and on the approximation that half of the galaxy total light is enclosed within a disk of radius equal to $R_{e}$, we estimate the values of the stellar mass within the effective radius $M_{*}(<R_{e})$ by halving the photometrically determined mean values of the total stellar mass $M_{*}$:
\begin{equation}
M_{*}(<R_{e})=M_{*} / 2.
\label{eq:03}
\end{equation}
The stellar masses are known with typical errors on the order of 20\%. 

Then, assuming that the total (luminous and dark) mass distribution is well described by a one-component isothermal distribution and that the value of $\sigma_{0}$ is representative of the value of the effective velocity dispersion of the isothermal model (for more details on the observational evidence relative to these points, see Sect. 1 and references therein), we measure the values of the total mass within the effective radius $M_{\mathrm{T}}(<R_{e})$ as follows:
\begin{equation}
M_{\mathrm{T}}(<R_{e})=\pi \sigma_{0}^{2} R_{e} / G.
\label{eq:04}
\end{equation}
The total masses have a typical error on the order of 12\%.

Next, we compute inside the effective radius the fraction of dark over total mass $f_{\mathrm{D}}(<R_{e})$ by subtracting to one the luminous over total mass fraction inside $R_{e}$:
\begin{equation}
f_{\mathrm{D}}(<R_{e})= 1 - M_{*}(<R_{e}) \,/\, M_{\mathrm{T}}(<R_{e})\,,
\label{eq:05}
\end{equation}
and the central surface dark matter density $\Sigma_{\mathrm{D},R_{e}}$ by dividing the estimate of the mass in the form of dark matter by the area of the disk delimited by $R_{e}$:
\begin{equation}
\Sigma_{\mathrm{D},R_{e}} =  \big[M_{\mathrm{T}}(<R_{e}) - M_{*}(<R_{e})\big] \,/\, \pi R_{e}^{2}.
\label{eq:06}
\end{equation}

\section{Results}

\begin{figure*}
\centering
\includegraphics[width=0.49\textwidth]{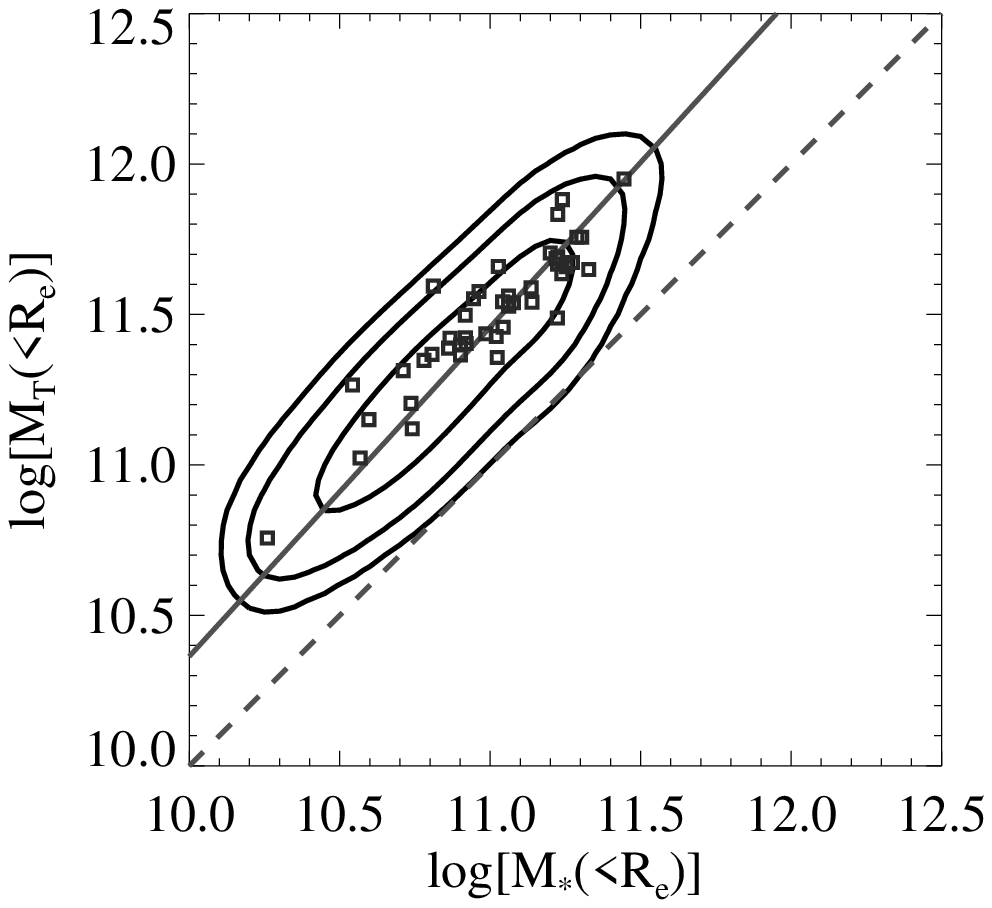}
\includegraphics[width=0.49\textwidth]{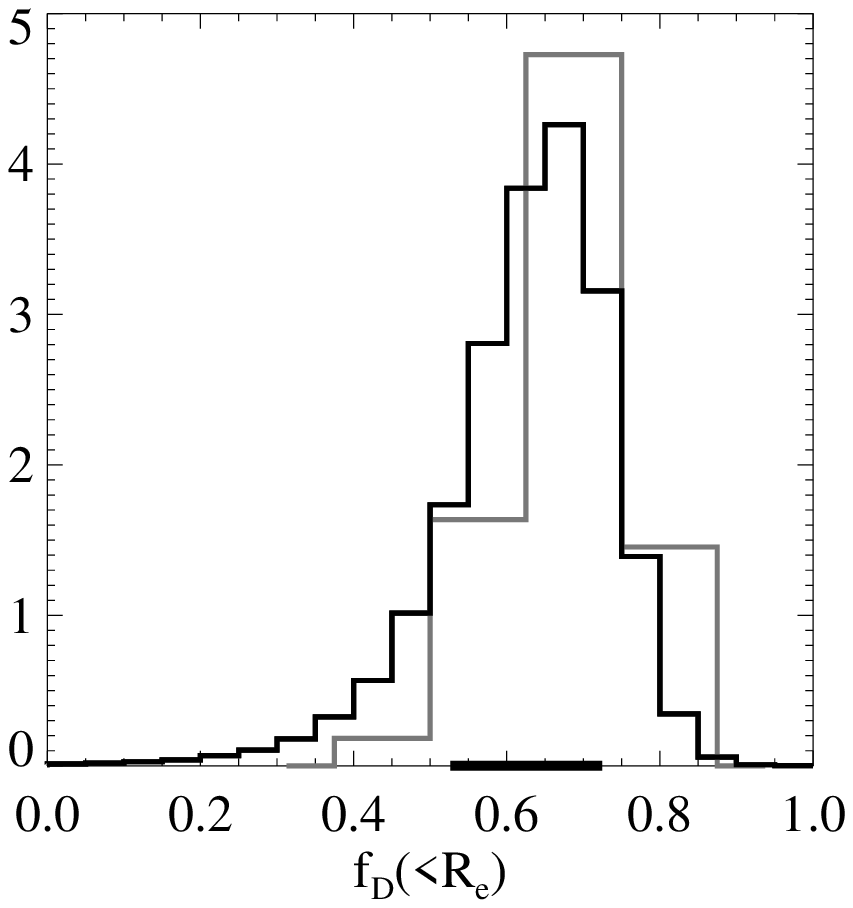}
\includegraphics[width=0.49\textwidth]{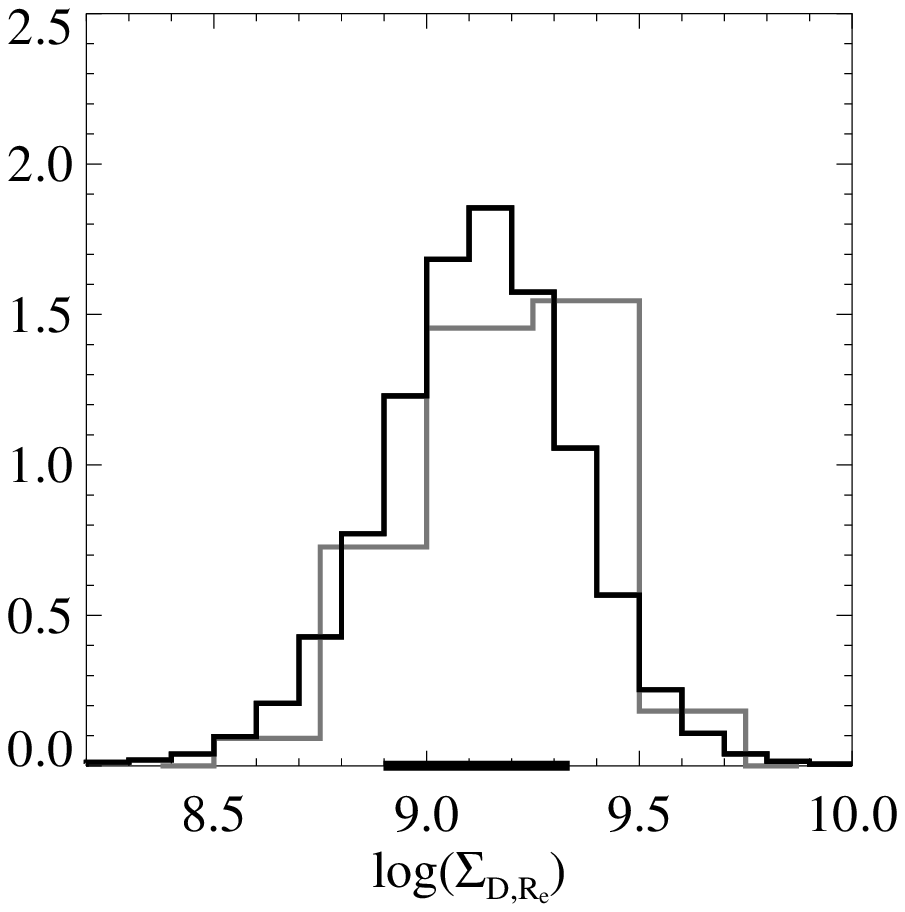}
\includegraphics[width=0.49\textwidth]{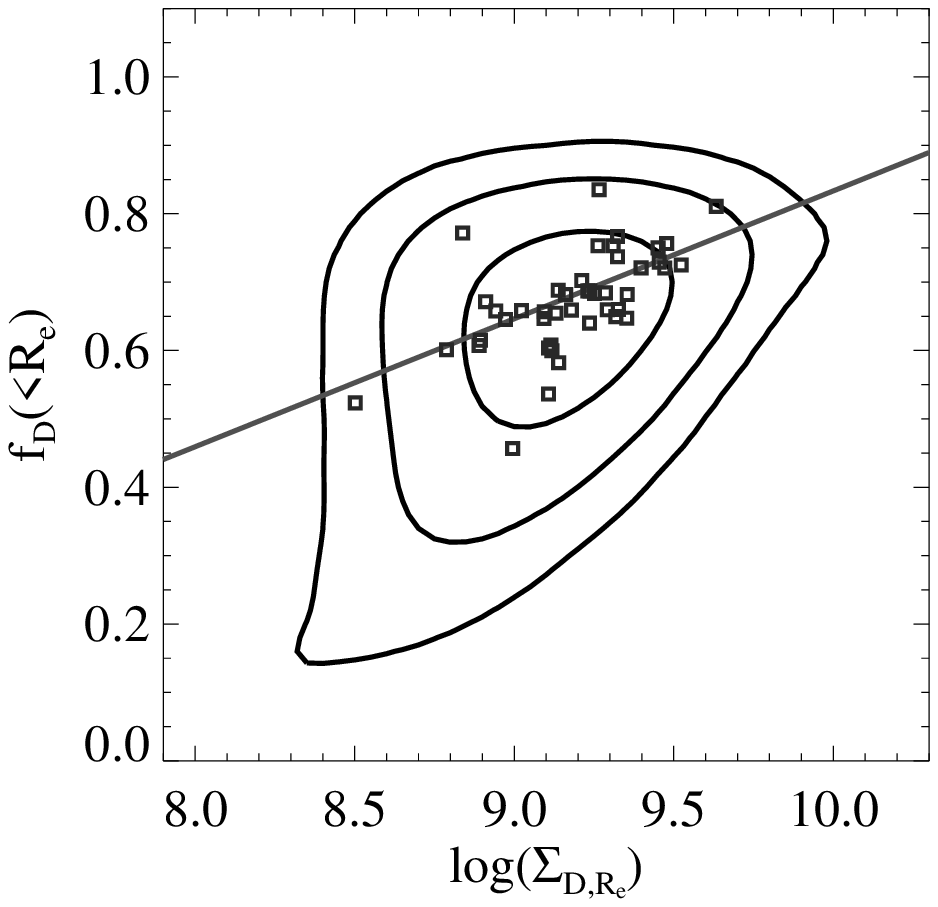}
\caption{\emph{On the top left:} The logarithmic values of the galaxy total $M_{\mathrm{T}}(<R_{e})$ and stellar $M_{*}(<R_{e})$ masses projected in a cylinder with radius equal to the effective radius $R_{e}$. \emph{On the top right:} The probability distribution function of the fractions of dark over total mass projected within the effective radius $f_{\mathrm{D}}(<R_{e})$. \emph{On the bottom left:} The probability distribution function of the logarithmic values of the dark matter density projected within the effective radius $\Sigma_{\mathrm{D},R_{e}}$. \emph{On the bottom right:} The values of the fraction of dark over total mass and the logarithmic values of the dark matter density projected in a cylinder with radius equal to the effective radius $R_{e}$. For the approximately $1.7 \times 10^{5}$ galaxies in the sample, the contour levels represent the 68\%, 95\%, and 99\% confidence level regions, and the solid and dashed lines show the best-fit and the one-to-one lines, respectively. The solid lines on the \emph{x}-axis represents the 68\% confidence intervals. The squares and the light color probability distribution functions indicate the 44 (grade-A) SLACS lens galaxies that satisfy the selection criteria of the sample.}
\label{fi02}
\end{figure*}

We plot in Fig. \ref{fi02} the 68\%, 95\%, and 99\% confidence regions of the luminous versus total masses and the dark matter fractions versus dark matter densities (all the quantities are estimated within $R_{e}$), and the probability distribution functions of the dark over total mass fractions and dark matter densities of all the galaxies in the sample. In each panel of Fig. \ref{fi02}, we also show the 44 lens galaxies of the SLACS survey that according to the adopted selection criteria belong to the sample studied here.

As expected, we find a highly statistically significant value of 0.91 for the Pearson linear correlation coefficient between the logarithmic values of $M_{\mathrm{T}}(<R_{e})$ and $M_{*}(<R_{e})$. The best-fit correlation line between the logarithmic values of the luminous and total masses of the galaxies in the sample is
\begin{equation}
\log \big[M_{\mathrm{T}}(<R_{e})\big] = -0.58 + 1.09 \times \log \big[M_{*}(<R_{e})\big].
\label{eq:07}
\end{equation}
The errors on the two best-fit linear coefficients are very small, given the large number of data points and their relatively small scatter around the best-fit line. Remarkably, if we compare the best-fit and one-to-one lines in Fig. \ref{fi02}, we can conclude that inside $R_{e}$ the galaxies in the sample show total masses that increase almost linearly with the luminous masses. This result is emphasized by looking at the second panel of Fig. \ref{fi02}. The dark over total mass fractions present a unimodal, slightly asymmetric, and almost Gaussian probability distribution, with a median value of 0.64 and 68\% confidence level values between 0.53 and 0.72. The logarithmic values of the dark matter density also exhibit a probability distribution function which is well approximated by a Gaussian function, with a median value of 9.1  and a 68\% confidence level interval extending from 8.9 to 9.3. We measure a Pearson linear correlation coefficient value of 0.49 between the values of the dark matter fraction and the logarithmic values of the dark matter density (for the best-fit line parameters, see Table \ref{ta01} and Fig. \ref{fi02}). This correlation suggests that an increasing fraction of dark over total mass within $R_{e}$ is associated to an increasing concentration inside the same radius of the dark matter profile.

The tight selection criteria adopted to define our galaxy sample guarantee the robustness of the previous conclusions. In fact, the requirement on the good approximation of the galaxy surface brightness distribution by a de Vaucouleurs profile ensures that the effective radius is suitable to represent the half-light radius. As a result, the value of the stellar mass within $R_{e}$ is well represented by half the value of the total stellar mass. In addition, the relatively small ranges of allowed values for the stellar velocity dispersion and stellar mass make certain that a one-component isothermal distribution is an appropriate parametrization of the total mass distribution. This last statement is supported by the results of numerous dynamical and lensing studies in smaller samples of galaxies that present structural properties similar to those of the galaxies in our sample (see Sect. 1).  

To further check how the previous result of an almost constant dark over total mass fraction is affected by some scatter in the one-to-one relation between the effective velocity dispersions of the one-component isothermal model and the central stellar velocity dispersions, we perform the following test. We extract new values of $\sigma_{0}$ from Gaussian probability distributions centered on the measured values and with standard deviations equal to 10\% of the measured values. Then, we use them to estimate the projected total masses and dark over total mass fractions, as described in the previous section. We find that the approximately linear relation observed in the first panel of Fig. \ref{fi02} between the logarithmic values of the luminous and total masses of the galaxies in the sample is conserved, despite some expected larger scatter. The probability distribution function of the dark over total mass fractions plotted in the second panel of Fig. \ref{fi02} is slightly broader, but not significantly modified. The median value is still 0.64 and the 68\% confidence level interval extends between 0.49 and 0.74. On the basis of this analysis, we confirm the robustness of our results. We remark that this test also ensures that small deviations from an average one-component isothermal model for the galaxy total mass distribution would act in the same way and not change appreciably the general picture presented so far.

We emphasize that the uncertainties on the total and stellar masses are not correlated. This ensures that the correlation pattern visible in Fig.~\ref{fi02} between $M_{\mathrm{T}}(<R_{e})$ and $M_{*}(<R_{e})$ is not affected by the error correlations. In fact, on the one hand, we have checked that the errors on the stellar masses are primarily attributable to the degeneracies that are inherent in the stellar population modeling and only to a secondary level to the photometric uncertainties. On the other hand, we observe that the errors on the total masses reflect essentially the precision with which the stellar velocity dispersions can be measured from the galaxy spectra.

In passing, we mention that the baryonic Tully-Fisher relation (e.g., \citealt{mcg00}), a scaling law between the total baryonic mass and the maximum rotation velocity of late-type galaxies, seems also to suggest that in these objects the dark over total mass fractions are constant over a wide mass range.

Recent analyses that combine photometric, dynamical, and lensing data in massive early-type galaxies seem to agree on favoring for these objects a Salpeter to a Chabrier or a Kroupa stellar IMF (e.g., \citealt{gri08a,gri09,gri10b,tre10}). According to these results, if we normalize in our sample the stellar mass estimates obtained by adopting a Chabrier IMF to a Salpeter IMF, we measure a median value of the dark over total matter fraction of approximately 0.4 within $R_{e}$. This result is consistent, given the errors, with the average values of 0.3 (see \citealt{koo06,gri09,aug09}) and 0.4 (see \citealt{bol08b}) for the 2D dark over total mass fractions determined, respectively, inside disks with apertures equal to the average Einstein (0.6 times the effective radius) and effective radii of the SLACS lens galaxies. Furthermore, in 6 of these galaxies \citet{bar09} have combined integral field spectroscopy with lensing data and measured values between 15\% and 30\% for the 3D dark over total mass fractions within $R_{e}$. Our results are consistent with these values and with the values between 20\% and 30\% for the same fractions that have been estimated by precise dynamical modeling of the photometric and spectroscopic observations of about 20 massive early-type galaxies of the Coma cluster (\citealt{tho07}). Unfortunately, a direct comparison with the values of the dark matter densities provided by \citet{tho09} and determined in the Coma galaxies with the same dynamical technique is not feasible because of projection effects and differently probed spatial regions. In fact, the dark matter densities of the Coma sample are measured within a sphere of radius equal to two times the value of $R_{e}$. 

\begin{table}
\centering
\caption{Best-fit parameters of the dark matter scaling relations ($y=a+b \times x$).}
\begin{tabular}{ccccc}
\hline\hline \noalign{\smallskip}
$y$ & $x$ & $a$ & $b$ & $\rho$ \\ 
\noalign{\smallskip} \hline
$f_{\mathrm{D}}(<R_{e})$ & $\log (M_{*})$ & 1.27 & $-$0.05 & $-$0.12 \\
$f_{\mathrm{D}}(<R_{e})$ & $\log (\sigma_{0})$ & $-$0.44 & 0.47 & 0.30 \\
$f_{\mathrm{D}}(<R_{e})$ & $\log (R_{e})$ & 0.60 & 0.11 & 0.18 \\
$f_{\mathrm{D}}(<R_{e})$ & $\log (\Sigma_{*,R_e})$ & 2.45 & $-$0.20 & $-$0.45 \\
$f_{\mathrm{D}}(<R_{e})$ & $\log (\Sigma_{\mathrm{D},R_e})$ & $-$1.04 & 0.19 & 0.46 \\ 
$\log (\Sigma_{\mathrm{D},R_e})$ & $\log (M_{*})$ & 11.8 & $-$0.24 & $-$0.29 \\
$\log (\Sigma_{\mathrm{D},R_e})$ & $\log (\sigma_{0})$ & 4.38 & 2.03 & 0.48 \\
$\log (\Sigma_{\mathrm{D},R_e})$ & $\log (R_{e})$ & 9.69 & $-$0.68 & $-$0.54 \\
$\log (\Sigma_{\mathrm{D},R_e})$ & $\log (\Sigma_{*,R_e})$ & 1.17 & 0.90 & 0.57 \\
\noalign{\smallskip} \hline
\end{tabular}
\label{ta01}
\end{table}

\begin{figure*}
\centering
\includegraphics[width=0.4\textwidth]{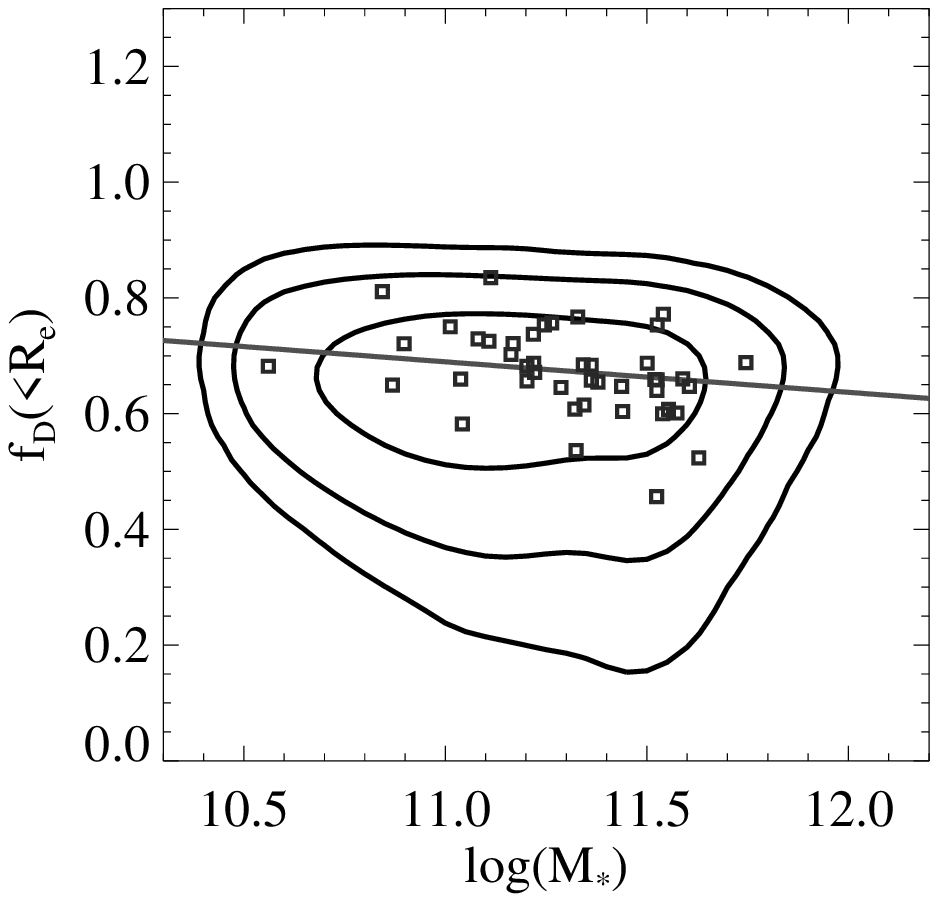}
\includegraphics[width=0.4\textwidth]{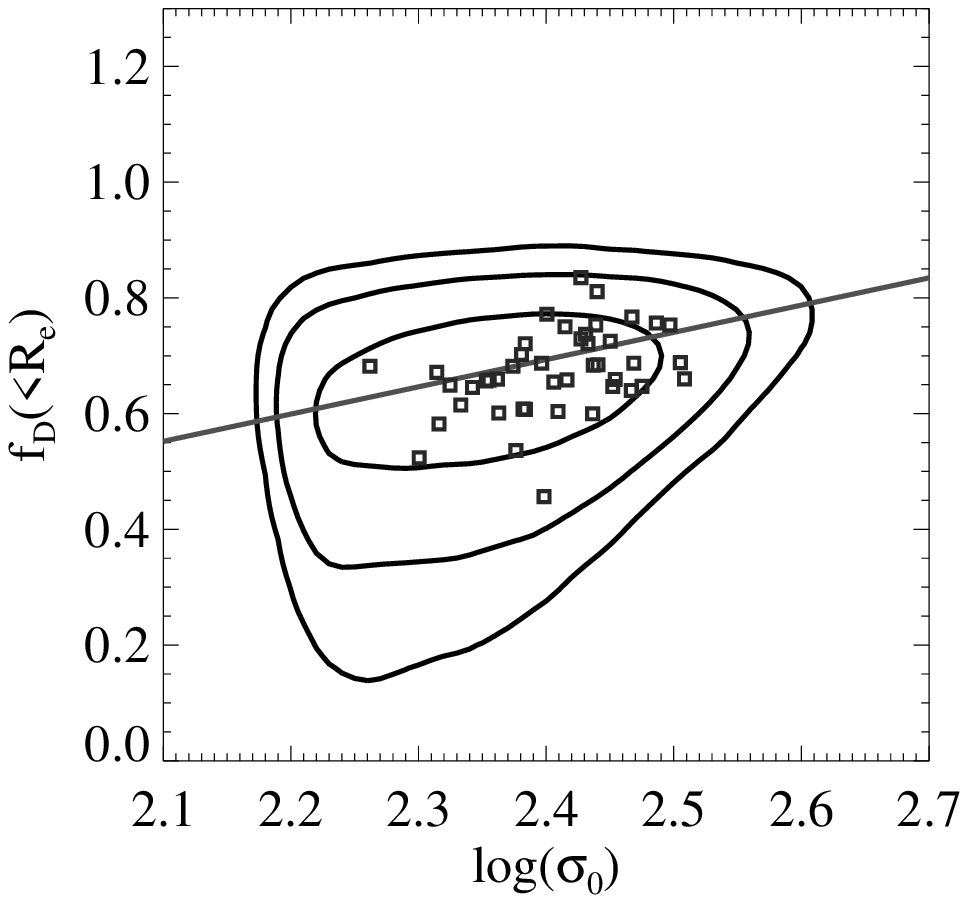}
\includegraphics[width=0.4\textwidth]{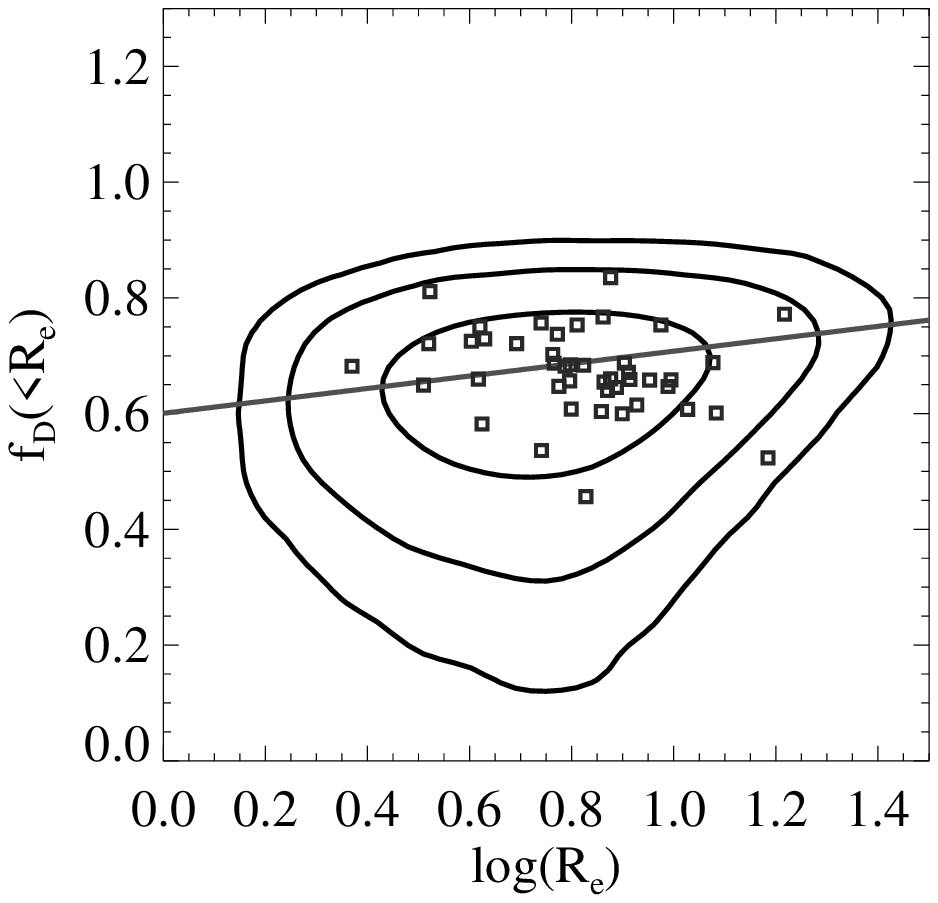}
\includegraphics[width=0.4\textwidth]{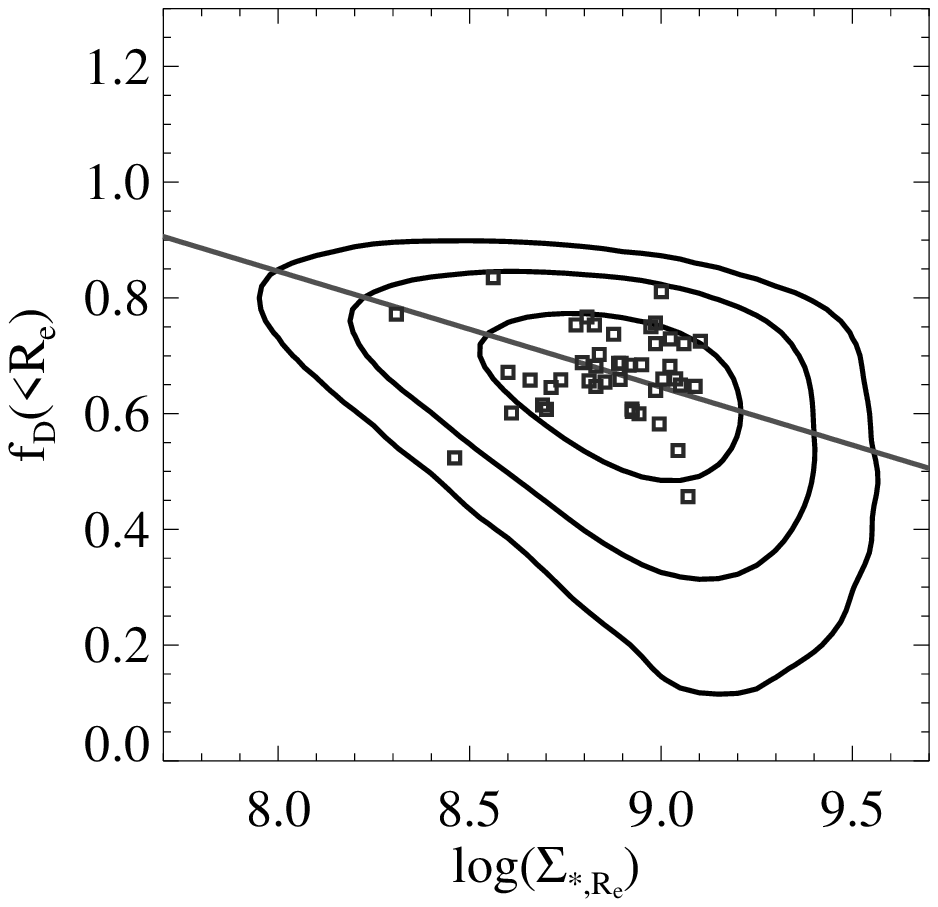}
\caption{The values of the fraction of dark over total mass projected within the effective radius $f_{\mathrm{D}}(<R_{e})$ as a function of the logarithmic values of the total stellar mass $M_{*}$ (\emph{on the top left}), central stellar velocity dispersion $\sigma_{0}$ (\emph{on the top right}), effective radius $R_{e}$ (\emph{on the bottom left}), and central surface stellar mass density $\Sigma_{*,R_{e}}$ (\emph{on the bottom right}) for the approximately $1.7 \times 10^{5}$ galaxies in the sample. The contour levels represent the 68\%, 95\%, and 99\% confidence level regions, and the solid lines show the best-fit lines. The squares indicate the 44 (grade-A) SLACS lens galaxies that satisfy the selection criteria of the sample.}
\label{fi03}
\end{figure*}

\begin{figure*}
\centering
\includegraphics[width=0.4\textwidth]{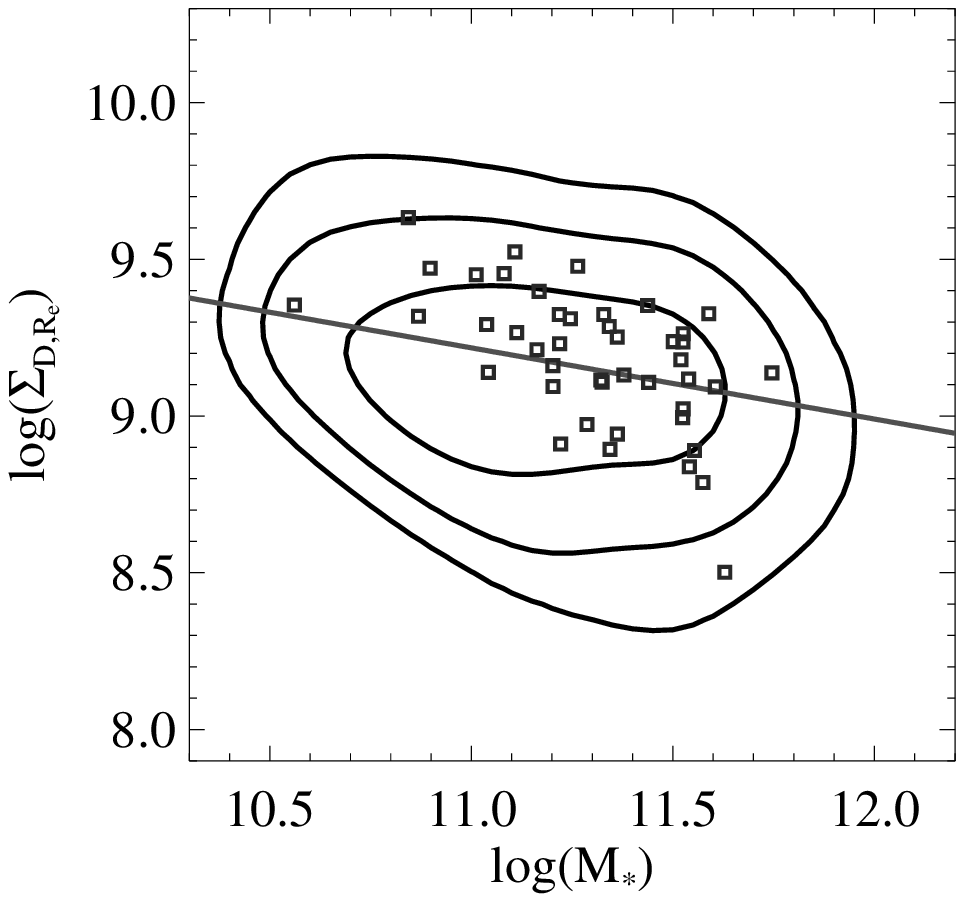}
\includegraphics[width=0.4\textwidth]{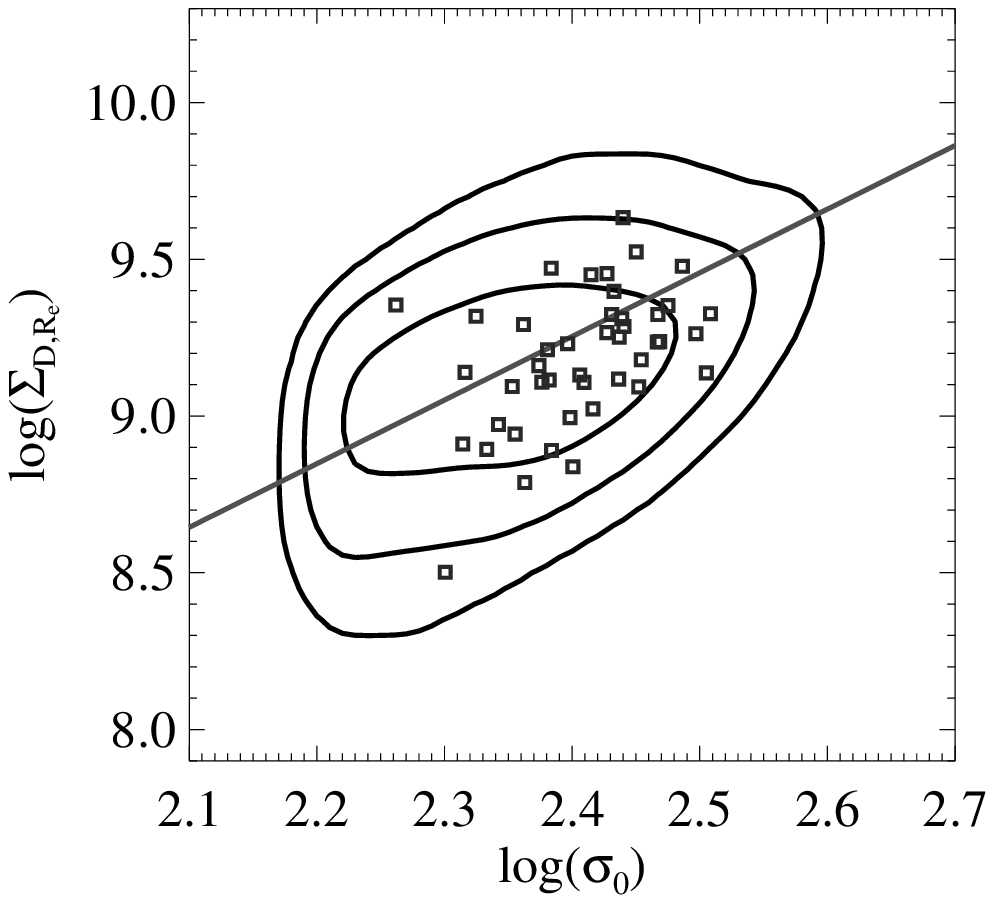}
\includegraphics[width=0.4\textwidth]{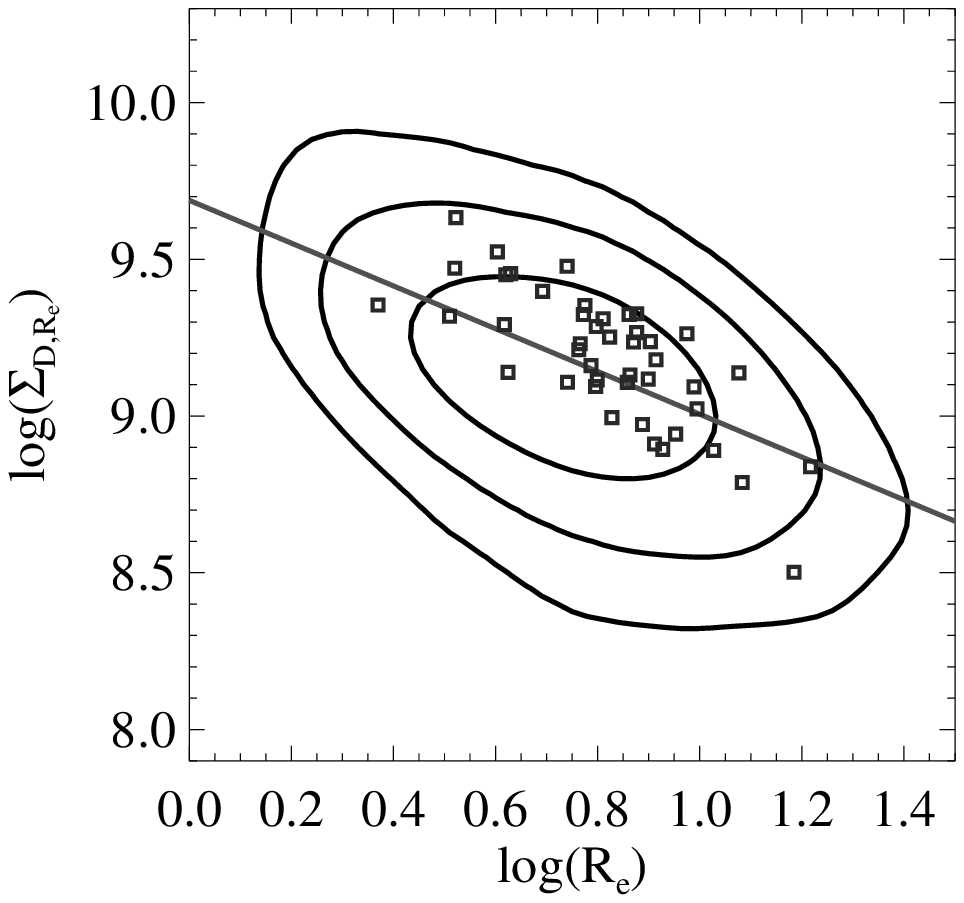}
\includegraphics[width=0.4\textwidth]{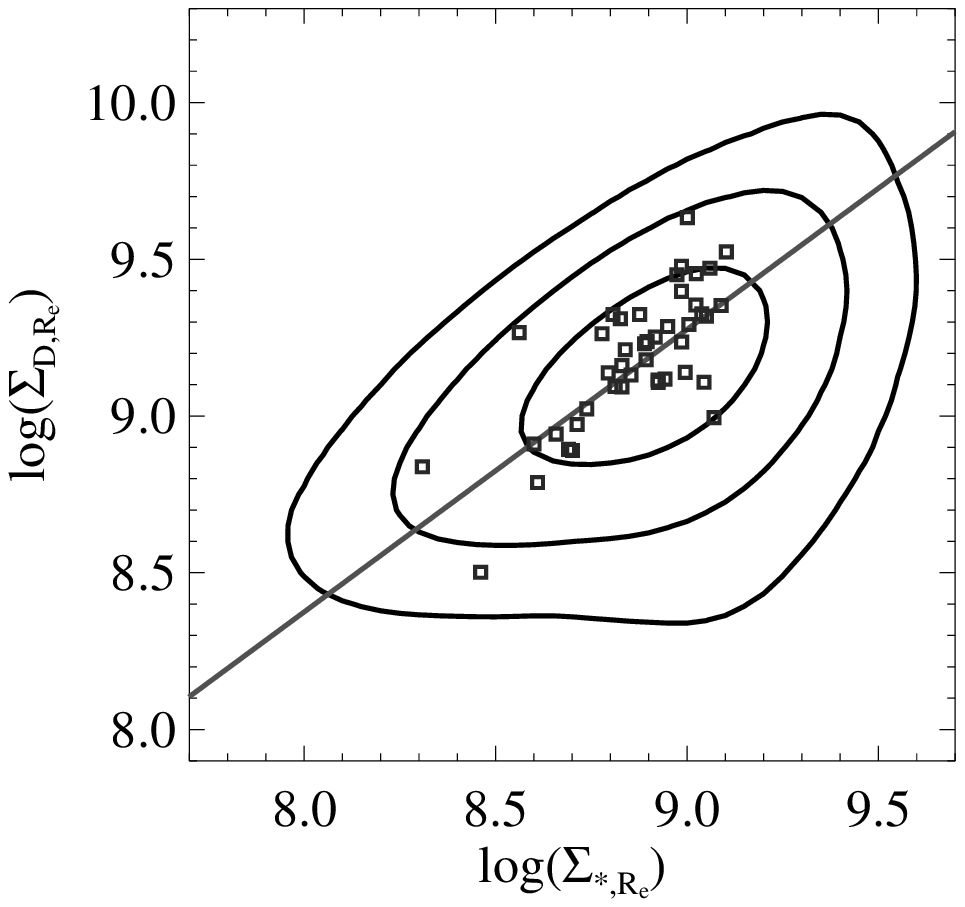}
\caption{The logarithmic values of the surface dark matter density projected within the effective radius $\Sigma_{\mathrm{D},R_{e}}$ as a function of the logarithmic values of the total stellar mass $M_{*}$ (\emph{on the top left}), central stellar velocity dispersion $\sigma_{0}$ (\emph{on the top right}), effective radius $R_{e}$ (\emph{on the bottom left}), and central surface stellar mass density $\Sigma_{*,R_{e}}$ (\emph{on the bottom right}) for the approximately $1.7 \times 10^{5}$ galaxies in the sample. The contour levels represent the 68\%, 95\%, and 99\% confidence level regions, and the solid lines show the best-fit lines. The squares indicate the 44 (grade-A) SLACS lens galaxies that satisfy the selection criteria of the sample.}
\label{fi04}
\end{figure*}

In Figs. \ref{fi03} and \ref{fi04} and Table \ref{ta01}, we show the relations, the best-fit lines, the best-fit linear parameters ($a$ and $b$), and the values of the Pearson linear correlation coefficient ($\rho$) between the dark over total mass fractions and the logarithmic values of the dark matter densities versus the logarithmic values of the total stellar mass, central stellar velocity dispersion, effective radius, and surface stellar mass density within the effective radius [$\Sigma_{*,R_{e}} = M_{*}(<R_{e})\,/\,\pi R_{e}^{2}$]. We observe that the values of $f_{\mathrm{D}}(<R_{e})$ are correlated at a higher significance level to the values of $\sigma_{0}$ and $\Sigma_{*,R_{e}}$ than to those of $M_{*}$ and $R_{e}$. The values of $\Sigma_{\mathrm{D},R_{e}}$ are correlated at a statistically significant level to the values of $\sigma_{0}$, $R_{e}$, and $\Sigma_{*,R_{e}}$ and at a low level to those of $M_{*}$. The almost linear relation between $\Sigma_{\mathrm{D},R_{e}}$ and $\Sigma_{*,R_{e}}$ is essentially another way of showing that within the effective radius both luminous and dark masses differ from the total mass by nearly constant proportionality factors (see also the first panel of Fig. \ref{fi01}).

We notice that in Figs. \ref{fi02}-\ref{fi04} the subsample of 44 massive lens galaxies of the SLACS survey represents well the general properties of the entire sample. We remark that the SLACS lens galaxies are to a first approximation selected according to the values of their velocity dispersion (see \citealt{bol06}). This justifies the presence in our plots of slightly more lenses in the regions where the velocity dispersions are high. Moreover, the probability that a galaxy produces multiple images of background sources increases if the galaxy central total mass density is high (e.g., \citealt{van09}; \citealt{man09}). This explains the observed moderate preference of the lenses to lie in regions characterized by high central mass densities.

Surprisingly, our previous result on a fairly constant dark over total mass fraction within the effective radius is at variance with some other studies (e.g., \citealt{pad04}; \citealt{tor09}). A first plausible explanation of this fact is that in these works the dark matter fractions of early-type galaxies are given within a three-dimensional radius and estimated by assuming some virial relation or using non-trivial dynamical models. Analyses of this kind are in general considerably more model-dependent than our study. Furthermore, the broader and more heterogeneous physical properties of some galaxy samples (often extending to significantly less massive galaxies) may be a possible source of confusion in the interpretation of the results. We have considered here only massive early-type galaxies that have been shown to be a fairly uniform class of galaxies in many respects, not least as far as their age and metallicity content is concerned (e.g., \citealt{gal05,gal06}).

\section{Discussion}

We reconsider here some of the relations presented in the previous section and explore the possible physical mechanisms that are at their origin and the implications in the framework of galaxy formation and evolution studies.

Early-type galaxies are found to occupy only a small fraction of the parameter space defined by the three variables effective radius, central stellar velocity dispersion, and effective surface brightness (e.g., \citealt{djo87}; \citealt{dre87}). The observed tight scaling law between these global properties, known as the FP, can be interpreted in terms of a systematic increase of galaxy effective (dynamical) mass-to-light ratio with effective (dynamical) mass (e.g., \citealt{fab87}; \citealt{ben92}; \citealt{van95}; \citealt{cio96}). This interpretation is usually referred to as the tilt of the FP relative to the expectation of the virial theorem. The origin of the tilt of the FP is still debated and it can be ascribed to variations in the stellar population properties, dark matter content, and/or structural properties of early-type galaxies (e.g., \citealt{hjo95,pru97,ber02,tru04,pad04,gri09,all09,tor09,gri10b}). 

Our measurement of an almost constant ratio between the luminous and total mass inside the effective radius excludes that the tilt of the FP of massive early-type galaxies is primarily due to variations in their dark matter fractions. This fact combined with the observed small deviations from homology in the mass-dynamical structure of massive early-type galaxies (e.g., \citealt{bol08b}, \citealt{cap06}) support the conclusions of different analyses on a main stellar population origin for the tilt of the FP of these objects (e.g., \citealt{gri09,all09,gri10b}). 

A few studies have claimed to have provided some observational evidence that the stellar IMF of early-type galaxies may not be universal and non-evolving (e.g., \citealt{van08}; \citealt{tre10}). If we followed these suggestions and explicitly allowed a change in the stellar IMF going from a Chabrier/Kroupa-like to a Salpeter-like IMF, moving from less to more massive galaxies in our sample, we would conclude that in the inner regions less massive galaxies are more dark-matter dominated than more massive ones. This is the opposite of what would be required to explain the tilt of the FP with varying the galaxy dark matter content. Alternatively, if we considered a variation in the stellar IMF for the galaxies in our sample, but fixed the previous result on an almost constant value of $f_{\mathrm{D}}(<R_{e})$, this would leave small place to only a very fine-tuned combination of variations in the stellar IMF and dark matter density profile. In particular, a change in the direction from a Chabrier/Kroupa to a Salpeter stellar IMF as described above, i.e., moving towards increasing galaxy masses, could only be well coupled with an increase in the concentration of the dark matter density distribution to preserve the dark over total mass fraction in the central regions.


N-body cosmological simulations have predicted that in an expanding universe cold dark matter particles collapse into self-similar halos with a diverging inner density profile (e.g., \citealt{nav96}; \citealt{moo99}). The more recent inclusion of the baryonic physics in simulations has shown that the luminous component can modify significantly the dark matter profile of a galaxy (e.g., \citealt{lac10} and references therein). The assembly of the stellar mass of an elliptical galaxy is not easy to simulate and involve different physical processes that can be simplistically classified as dissipationless and dissipational, depending on if the total energy of the system is conserved or not. Two possible formation scenarios are one in which stars originally form far from the effective radius of a galaxy and are subsequently accreted in its inner regions, and the other in which gas first flows into the central regions of a halo and is then transformed, once in place, into stars. In the first case, the stellar clumps loose their orbital energy through dynamical friction by heating the dark matter halo and hence smoothing its inner density cusp (e.g., \citealt{elz01}; \citealt{ber03}; \citealt{ma04}). In the second case, the gas radiates away its orbital and thermal energy and is responsible for the adiabatic contraction of the dark matter halo, i.e., for a central density profile steeper with respect to the initial distribution (e.g., \citealt{blu86}; \citealt{jes02}; \citealt{gne04}). In realistic models, both processes are expected to occur and their relative contribution determines the actual distribution of dark matter in the centre of a halo. For massive ellipticals, smoothed particle hydrodynamical simulations (e.g., \citealt{naa07}; \citealt{joh09}) seem to favor a star formation picture which is characterized by an initial dissipative collapse of gas with quick ``in -situ'' star formation (at redshift $\gtrsim 2$), followed by an extended phase of significant accretion of stellar material (at redshift $\lesssim 3$). Following the simulated time evolution of the stellar mass assembly in a massive galaxy, it is possible to notice that a remarkable fraction of the stars observed at the effective radius at redshift $z \simeq 0$ was not born there, but has been captured and accumulated in this region. The accretion of these stars, also called dry (gas-poor) minor mergers, is thought to be very important for the structural evolution of massive early-type galaxies and, in particular, at the origin of the late size, mass, and density evolution of early-type galaxies (e.g., \citealt{naa09}).

The correlations observed in the first and third panels of Fig. \ref{fi04} between the projected dark matter density and stellar mass and effective radius may be connected to the discussed mechanism of dissipationless accretion of groups of stars formed outside the galaxy central regions. According to this picture, galaxies with a large number of captured satellites of stars should on average show light distribution that are more diffused, hence larger values of the effective radii, compared to galaxies with less stellar satellites. Moreover, galaxies with high luminous mass may be those that have been able to accrete more external stellar systems. A large number of stellar clumps located in the inner regions of the galaxies with the largest values of effective radius and luminous mass can have smoothed the central cusp of the dark matter density profile appreciably. As a consequence, in the dissipationless scenario of galaxy evolution an anti-correlation between the values of the central (surface) dark matter density and effective radius and stellar mass is likely to be expected.  

Other processes which can contribute to determine the amount and concentration of dark matter in the central regions of galaxies are feedback from supernovae and active galactic nuclei (AGN). The latter is thought to be particularly important in galaxies similar to those in our sample, i.e., characterized by high velocity dispersions (e.g., \citealt{sil98}; \citealt{cio01}). In detail, feedback from AGN is believed to be able to heat the cold gas in the central regions of massive halos and quench star formation there (e.g., \citealt{spr05}; \citealt{saz05}; \citealt{cio07}). Despite the overall good agreement on this last point, it is still debated which mode of AGN activity is most important. A merger-induced ``quasar mode'' is thought to lead to an initial starburst followed by a suppression of star formation and an expulsion of the gas from the galaxy center, once the super massive black holes become sufficiently massive. Given the extensively studied relation between the mass of a black hole and the stellar velocity dispersion of its host galaxy (e.g., \citealt{fer00}; \citealt{geb00}), the effect of this mechanism on the dark matter distribution should be more pronounced in galaxies which exhibit large values of stellar velocity dispersion. Furthermore, the findings of several other studies (e.g., \citealt{ric98}; \citealt{kor09}) that have focused on the orbital decay of binary massive black holes arising from dissipationless mergers of gas-poor galaxies seem to predict results that point to the same direction. In fact, this process may be at the origin of the formation of the observed cores in the surface brightness distribution of high-luminosity early-type galaxies. The scouring of a core in the stellar mass distribution of a galaxy through the ejection of stars from the galaxy center by the orbiting of the black hole binary is expected to depend on the mass of the ultimately coalesced black hole, hence, on the value of the stellar velocity dispersion of the remnant galaxy.

The second panels of Figs. \ref{fi03} and \ref{fi04} do indeed show that galaxies with higher central stellar velocity dispersions present larger fractions and higher (surface) mass densities of dark matter projected within the effective radius. We then speculate that AGN feedback and/or massive black hole merging may contribute to the origin of these correlations.

We notice that the measurements obtained in this work offer the interesting opportunity to test the predictions of numerical simulations and to quantify the relevance of the different physical processes that are believed to play a role in galaxy formation and evolution.






\section{Conclusions}

In this paper we have selected a sample of nearly $1.7 \times 10^{5}$ massive early-type galaxies at redshift smaller than 0.33 from the SDSS DR7 to study their projected dark over total mass fraction and surface dark matter density within the effective radius. We have obtained from the MPA/JHU galaxy catalog the stellar mass values, that were estimated by fitting the SDSS multi-band photometry with a large grid of composite stellar population models and adopting a Chabrier stellar IMF. By generalizing the results derived from gravitational lensing and stellar dynamics analyses on a subsample of lens galaxies, we have modeled the total mass distribution of the galaxies in the sample with a one-component isothermal model with effective velocity dispersion equal to the central stellar velocity dispersion and, hence, measured the galaxy total mass values inside their effective radii. Then, we have combined the stellar and total mass measurements to determine the amount and concentration of dark matter within the galaxy half-light radii.    
Our main results can be summarized in the following points:
\begin{itemize}

\item[$-$] The projected total mass values are to a very good approximation linearly proportional to the stellar ones. This implies an almost constant fraction of dark over total mass contained within the galaxy effective radii. In detail, the dark matter component accounts on average for approximately 60\% of the two-dimensional total masses.

\item[$-$] The average logarithmic value of the surface dark matter density (in units of $M_{\odot}\, \mathrm{kpc}^{-2}$) inside $R_{e}$ is 9.1. The galaxies with the largest fractions of dark over total mass have the largest surface dark matter densities within the effective radii.

\item[$-$] If the stellar IMF and the dark matter density profile do not show significant variations with the total stellar mass of the sample galaxies, the tilt of the FP of massive early-type galaxies cannot be primarily ascribed to variations in their central dark matter content.

\item[$-$] The observed correlations of the values of $f_{\mathrm{D}}(<R_{e})$ and $\Sigma_{\mathrm{D},R_{e}}$ with those of $M_{*}$, $\sigma_{0}$, $R_{e}$, and $\Sigma_{*,R_{e}}$ provide positive evidence on the importance of dry minor mergers, AGN feedback, and/or binary black hole coalescence in driving the structural evolution of massive early-type galaxies. 

\end{itemize}
  
We remark that a plausible and viable solution to a better understanding of the results presented in this study awaits for cosmological simulations that are able to model the physics of baryons at a more accurate level than reached so far. The reproduction through simulation of these observational findings would provide an invaluable step forward in our insight into the main mechanisms that rule the initial formation and the subsequent evolution of massive early-type galaxies. 





\acknowledgments

C. G. is grateful to Marco Lombardi, Piero Rosati, and Stella Seitz for interesting talks and useful suggestions. This research was supported by the DFG cluster of excellence ``Origin and Structure of the Universe''. 

This work has made extensive use of the SDSS database. Funding for the SDSS and SDSS-II has been provided by the Alfred P. Sloan Foundation, the Participating Institutions, the National Science Foundation, the U.S. Department of Energy, the National Aeronautics and Space Administration, the Japanese Monbukagakusho, the Max Planck Society, and the Higher Education Funding Council for England. The SDSS Web Site is http://www.sdss.org/. The SDSS is managed by the Astrophysical Research Consortium for the Participating Institutions. The Participating Institutions are the American Museum of Natural History, Astrophysical Institute Potsdam, University of Basel, University of Cambridge, Case Western Reserve University, University of Chicago, Drexel University, Fermilab, the Institute for Advanced Study, the Japan Participation Group, Johns Hopkins University, the Joint Institute for Nuclear Astrophysics, the Kavli Institute for Particle Astrophysics and Cosmology, the Korean Scientist Group, the Chinese Academy of Sciences (LAMOST), Los Alamos National Laboratory, the Max-Planck-Institute for Astronomy (MPIA), the Max-Planck-Institute for Astrophysics (MPA), New Mexico State University, Ohio State University, University of Pittsburgh, University of Portsmouth, Princeton University, the United States Naval Observatory, and the University of Washington.






\clearpage




\clearpage

\clearpage


\begin{thebibliography}{}

\bibitem[Allanson et al.(2009)]{all09} Allanson, S. P., Hudson, M. J., Smith, R. J., and Lucey, J. R. 2009, \apj, 702, 1275
\bibitem[Arnaboldi et al.(1998)]{arn98} Arnaboldi, M., Freeman, K. C., Gerhard, O., Matthias, M., Kudritzki, R. P., M\'endez, R. H., Capaccioli, M., and Ford, H. 1998, \apj, 507, 759
\bibitem[Auger et al.(2009)]{aug09} Auger, M. W., Treu, T., Bolton, A. S., Gavazzi, R., Koopmans, L. V. E., Marshall, P. J., Bundy, K., and Moustakas, L. A. 2009, \apj, 705, 1099
\bibitem[Barnab\`e et al.(2009)]{bar09} Barnab\`e, M., Czoske, O., Koopmans, L. V. E., Treu, T., Bolton, A., and Gavazzi, R. 2009, \mnras, 399, 21 
\bibitem[Bender et al.(1992)]{ben92} Bender, R., Burstein, D., \& Faber, S. M. 1992, \apj, 399, 462 
\bibitem[Bertin et al.(1992)]{ber92} Bertin, G., Saglia, R. P., \& Stiavelli, M. 1992, \apj, 384, 423
\bibitem[Bertin et al.(1994)]{ber94} Bertin, G., Bertola, F., Buson, L. M., Danzinger, I. J., Dejonghe, H., Sadler, E. M., Saglia, R. P., de Zeeuw, P. T., and Zeilinger, W. W. 1994, \aap, 292, 381
\bibitem[Bertin et al.(2002)]{ber02} Bertin, G., Ciotti, L., Del Principe, M. 2002, \aap, 386, 149
\bibitem[Bertin et al.(2003)]{ber03} Bertin, G., Liseikina, T., \& Pegoraro, F. 2003, \aap, 405, 73
\bibitem[Blumenthal et al.(1986)]{blu86} Blumenthal, G. R., Faber, S. M., Flores, R., and Primack, J. R. 1986, \apj, 301, 27
\bibitem[Bolton et al.(2006)]{bol06} Bolton, A. S., Burles, S., Koopmans, L. V. E., Treu, T., and Moustakas, L. A. 2006, \apj, 638, 703
\bibitem[Bolton et al.(2008a)]{bol08a} Bolton, A. S., Burles, S., Koopmans, L. V. E., Treu, T., Gavazzi, R., Moustakas, L. A., Wayth, R., and Schlegel, D. J. 2008a, \apj, 682, 964
\bibitem[Bolton et al.(2008b)]{bol08b} Bolton, A. S., Treu, T., Koopmans, L. V. E., Gavazzi, R., Moustakas, L. A., Burles, S., Schlegel, D. J., and Wayth, R. 2008b, \apj, 684, 248
\bibitem[Bruzual \& Charlot(2003)]{bru03} Bruzual, G. \& Charlot, S. 2003, MNRAS, 344, 1000
\bibitem[Buson et al.(1993)]{bus93} Buson, L. M., et al. 1993, \aap, 280, 409
\bibitem[Cappellari et al.(2006)]{cap06} Cappellari, M., et al. 2006, MNRAS, 366, 1126
\bibitem[Carollo et al.(1995)]{car95} Carollo, C. M., de Zeeuw, P. T., van der Marel, R. P., Danziger, I. J., and Qian, E. E. 1995, \apj, 441, 25
\bibitem[Chabrier(2003)]{cha03} Chabrier, G. 2003, PASP, 115, 763
\bibitem[Ciotti et al.(1996)]{cio96} Ciotti, L., Lanzoni, B., \& Renzini, A. 1996, \mnras, 282, 1
\bibitem[Ciotti \& Ostriker(2001)]{cio01} Ciotti, L. \& Ostriker, J. P. 2001, \apj, 551, 131
\bibitem[Ciotti \& Ostriker(2007)]{cio07} Ciotti, L. \& Ostriker, J. P. 2007, \apj, 665, 1038
\bibitem[Czoske et al.(2008)]{czo08} Czoske, O., Barnab\`e, M., Koopmans, L. V. E., Treu, T., and Bolton, A. S. 2008, \mnras, 384, 987
\bibitem[Djorgovski \& Davies(1987)]{djo87} Djorgovski, S. \& Davies, M. 1987, \apj, 313, 59
\bibitem[Dressler et al.(1987)]{dre87} Dressler, A., Lynden-Bell, D., Burstein, D., Davies, R. L., Faber, S. M., Terlevich, R., and Wegner, G. 1987, \apj, 313, 42
\bibitem[El-Zant et al.(2001)]{elz01} El-Zant, A., Shlosman, I., \& Hoffman, Y. 2001, \apj, 560, 636
\bibitem[Faber et al.(1987)]{fab87} Faber, S. M., Dressler, A., Davies, R. L., Burstein, D., and Lynden-Bell, D. 1987, in Nearly Normal Galaxies, ed. S. M. Faber (New York: Springer), 175
\bibitem[Fadely et al.(2010)]{fad10} Fadely, R., Keeton, C. R., Nakajima, R., and Bernstein, G. M. 2010, \apj, 711, 246
\bibitem[Falco et al.(1985)]{fal85} Falco, E. E., Gorenstein, M. V., \& Shapiro, I. I. 1985, \apj, 289, 1
\bibitem[Ferrarese \& Merritt(2000)]{fer00} Ferrarese, L. \& Merritt, D. 2000, \apj, 539, 9
\bibitem[Franx et al.(1994)]{fra94} Franx, M., van Gorkom, J. H., \& de Zeeuw, T. 1994, ApJ, 436, 642
\bibitem[Fukugita et al.(1996)]{fuk96} Fukugita, M., Ichikawa, T., Gunn, J. E., Shimasaku, K., and Schneider, D. P. 1996, \aj, 111, 1748
\bibitem[Gallazzi et al.(2005)]{gal05} Gallazzi, A., Charlot, S., Brinchmann, J., White, S. D. M., and Tremonti, C. A. 2005, \mnras, 362, 41
\bibitem[Gallazzi et al.(2006)]{gal06} Gallazzi, A., Charlot, S., Brinchmann, J., and White, S. D. M. 2006, \mnras, 370, 1106
\bibitem[Gavazzi et al.(2007)]{gav07} Gavazzi, R., Treu, T., Rhodes, J. D., Koopmans, L. V. E., Bolton, A. S., Burles, S., Massey, R. J., and Moustakas, L. A. 2007, \apj, 667, 176
\bibitem[Gebhardt et al.(2000)]{geb00} Gebhardt K., et al. 2000, \apj, 539, 13
\bibitem[Gerhard(1993)]{ger93} Gerhard, O. 1993, \mnras, 265, 213
\bibitem[Gerhard et al.(2001)]{ger01} Gerhard, O., Kronawitter, A., Saglia, R. P., and Bender, R. 2001, AJ, 121, 1936
\bibitem[Gnedin et al.(2004)]{gne04} Gnedin, O. Y., Kravtsov, A. V., Klypin, A. A., and Nagai, D. 2004, \apj, 616, 16
\bibitem[Grillo et al.(2008a)]{gri08a} Grillo, C., Gobat, R., Rosati, P., and Lombardi, M. 2008a, \aap, 477, 25
\bibitem[Grillo et al.(2008b)]{gri08b} Grillo, C., Lombardi, M., \& Bertin, G. 2008b, \aap, 477, 397
\bibitem[Grillo et al.(2008c)]{gri08c} Grillo, C., et al. 2008c, \aap, 486, 45
\bibitem[Grillo et al.(2009)]{gri09} Grillo, C., Gobat, R., Lombardi, M., and Rosati, P. 2009, \aap, 501, 461
\bibitem[Grillo et al.(2010a)]{gri10a} Grillo, C., Eichner, T., Seitz, S., Bender, R., Lombardi, M., Gobat, R., and Bauer, A. 2010a, \apj, 710, 372
\bibitem[Grillo \& Gobat(2010b)]{gri10b} Grillo, C. \& Gobat, R. 2010b, \mnras, 402, 67
\bibitem[Hernquist(1990)]{her90} Hernquist, L. 1990, \apj, 356, 359
\bibitem[Hjorth \& Madsen(1995)]{hjo95} Hjorth, J. \& Madsen, J. 1995, \apj, 445, 55
\bibitem[Jaffe(1983)]{jaf83} Jaffe, W. 1983, \mnras, 202, 995
\bibitem[Jesseit et al.(2002)]{jes02} Jesseit, R., Naab, T., \& Burkert, A. 2002, \apj, 571, 89
\bibitem[Johansson et al.(2009)]{joh09} Johansson, P. H., Naab, T., \& Ostriker, J. P. 2009, \apj, 697, 38
\bibitem[J\o rgensen et al.(1995)]{jor95} J\o rgensen, I., Franx, M., \& Kj\ae rgaard, P. 1995, \mnras, 276, 1341
\bibitem[Kauffmann et al.(2003)]{kau03} Kauffmann, G., et al. 2003, \mnras, 341, 33
\bibitem[Kochanek(1991)]{koc91} Kochanek, C. S. 1991, \apj, 373, 354
\bibitem[Kochanek(1993)]{koc93} Kochanek, C. S. 1993, \apj, 419, 12
\bibitem[Kochanek(1994)]{koc94} Kochanek, C. S. 1994, \apj, 436, 56
\bibitem[Kochanek(2006)]{koc06} Kochanek, C. S. 2006, Gravitational lensing: strong, weak and micro, Saas-Fee Advanced Course 33, ed. G. Meylan, P. Jetzer, and P. North (Berlin: Springer), 91-268
\bibitem[Komatsu et al.(2009)]{kom09} Komatsu, E., et al. 2009, ApJS, 180, 330
\bibitem[Koopmans et al.(2006)]{koo06} Koopmans, L. V. E., Treu, T., Bolton, A. S., Burles, S., and Moustakas, L. A. 2006, \apj, 649, 599
\bibitem[Koopmans et al.(2009)]{koo09} Koopmans, L. V. E., et al. 2009, \apj, 703, 51
\bibitem[Kormendy \& Bender(2009)]{kor09} Kormendy, J. \& Bender, R. 2009, \apj, 691, 142
\bibitem[Lackner \& Ostriker(2010)]{lac10} Lackner, C. N. \& Ostriker, J. P. 2010, \apj, 712, 88
\bibitem[Loewenstein \& White(1999)]{loe99} Loewenstein, M. \& White, R. E. 1999, \apj, 518, 50
\bibitem[\L okas \& Mamon(2003)]{lok03} \L okas, E. L. \& Mamon, G. A. 2003, \mnras, 343, 401
\bibitem[Ma \& Boylan-Kolchin(2004)]{ma04} Ma, C.-P. \& Boylan-Kolchin, M. 2004, PhRvL, 93, 1301
\bibitem[Mandelbaum et al.(2009)]{man09} Mandelbaum, R., van de Ven, G., \& Keeton, C. R. 2009, \mnras, 398, 635
\bibitem[McGaugh et al.(2000)]{mcg00} McGaugh, S. S., Schombert, J. M., Bothun, G. D., and de Blok, W. J. G. 2000, \apj, 533, 99
\bibitem[Moore et al.(1999)]{moo99} Moore, B., Quinn, T., Governato, F., Stadel, J., and Lake, G. 1999, \mnras, 310, 1147 
\bibitem[Mould et al.(1990)]{mou90} Mould, J. R., Oke, J. B., de Zeeuw, P. T., and Nemec, J. M. 1990, AJ, 99, 1823
\bibitem[Mushotzky et al.(1994)]{mus94} Mushotzky, R. F., Loewenstein, M., Awaki, H., Makishima, K., Matsushita, K., and Matsumoto, H. 1994, \apj, 436, 79
\bibitem[Naab et al.(2007)]{naa07} Naab, T., Johansson, P. H., Ostriker, J. P., and Efstathiou, G. 2007, \apj, 658, 710
\bibitem[Naab et al.(2009)]{naa09} Naab, T., Johansson, P. H., \& Ostriker, J. P. 2009, \apj, 699, 178
\bibitem[Navarro et al.(1996)]{nav96} Navarro, J., Frenk, C. S., \& White, S. D. M. 1996, \apj, 462, 563
\bibitem[Padmanabhan et al.(2004)]{pad04} Padmanabhan, N., et al. 2004, NewA, 9, 329
\bibitem[Prugniel \& Simien(1997)]{pru97} Prugniel, P. \& Simien, F. 1997, \aap, 321, 111
\bibitem[Richstone et al.(1998)]{ric98} Richstone, E. A., et al. 1998, Natur., 395, 14
\bibitem[Rusin et al.(2003)]{rus03} Rusin, D., Kochanek, C. S., \& Keeton, C. R. 2003, \apj, 595, 29
\bibitem[Saglia et al.(1992)]{sag92} Saglia, R. P., Bertin, G. \& Stiavelli, M. 1992, \apj, 384, 433
\bibitem[Salim et al.(2007)]{sal07} Salim, S., et al. 2007, \apjs, 173, 267  
\bibitem[Sazonov et al.(2005)]{saz05} Sazonov, S. Y., Ostriker, J. P., Ciotti, L., Sunyaev, R. A. 2005, \mnras, 358, 168
\bibitem[Silk \& Rees(1998)]{sil98} Silk, J. \& Rees, M. J. 1998, \aap, 331, 1
\bibitem[Springel et al.(2005)]{spr05} Springel, V., Di Matteo, T., \& Hernquist, L. 2005, \apj, 620, 79
\bibitem[Springel et al.(2006)]{spr06} Springel, V., Frenk, C. S., \& White, S. D. M. 2006, Natur., 440, 1137
\bibitem[Strauss et al.(2002)]{str02} Strauss, M. A., et al. 2002, \aj, 124, 1810
\bibitem[Thomas et al.(2007)]{tho07} Thomas, J., Saglia, R. P., Bender, R., Thomas, D., Gebhardt, K., Magorrian, J., Corsini, E. M., and Wegner, G. 2007, \mnras, 382, 657
\bibitem[Thomas et al.(2009)]{tho09} Thomas, J., Saglia, R. P., Bender, R., Thomas, D., Gebhardt, K., Magorrian, J., Corsini, E. M., and Wegner, G. 2009, \apj, 691, 770
\bibitem[Tortora et al.(2009)]{tor09} Tortora, C., Napolitano, N. R., Romanowsky, A. J., Capaccioli, M., and Covone, G. 2009, \mnras, 396, 1132
\bibitem[Treu \& Koopmans(2004)]{tre04} Treu, T. \& Koopmans, L. V. E. 2004, \aj, 611, 739 
\bibitem[Treu et al.(2006)]{tre06} Treu, T., Koopmans, L. V. E., Bolton, A. S., Burles, S., and Moustakas, L. A. 2006, \apj, 640, 662
\bibitem[Treu et al.(2009)]{tre09} Treu, T., Gavazzi, R., Gorecki, A., Marshall, P. J., Koopmans, L. V. E., Bolton, A. S., Moustakas, L. A., and Burles, S. 2009, \apj, 690, 670
\bibitem[Treu et al.(2010)]{tre10} Treu, T., Auger, M. W., Koopmans, L. V. E., Gavazzi, R., Marshall, P. J., and Bolton, A. S. 2010, \apj, 709, 1195
\bibitem[Trujillo et al.(2004)]{tru04} Trujillo, I., Burket, A., \& Bell, E. 2004, \apj, 600, 39
\bibitem[van Albada et al.(1995)]{van95} van Albada, T. S., Bertin, G., \& Stiavelli, M. 1995, \mnras, 276, 1255
\bibitem[van de Ven et al.(2009)]{van09} van de Ven, G., Mandelbaum, R., \& Keeton, C. R. 2009, \mnras, 398, 607
\bibitem[van der Marel(1991)]{van91} van Der Marel, R. P. 1991, \mnras, 253, 710
\bibitem[van Dokkum(2008)]{van08} van Dokkum, P. G. 2008, \apj, 674, 29
\bibitem[Vegetti et al.(2010)]{veg10} Vegetti, S., Koopmans, L. V. E., Bolton, A., Treu, T., and Gavazzi, R. 2010, arXiv:0910.0760
\bibitem[York(1966)]{yor66} York, D. 1966, Can. J. Phys., 44, 1079  

\end{thebibliography}
\end{document}